# Waveform-Logmel Audio Neural Networks for Respiratory Sound Classification

Jiadong Xie
*College of Integrated Circuits, State Key Laboratory of Silicon and Advanced Semiconductor Mateials*
*Zhejiang University*
Hangzhou, China
jonwitdo04@gmail.com

Yunlian Zhou*
*Department of Pulmonology, Children's Hospital*
*Zhejiang University*
Hangzhou, China
zhouyunlian@zju.edu.cn

Mingsheng Xu*
*College of Integrated Circuits*
*Zhejiang University*
Hangzhou, China
msxu@zju.edu.cn

*Abstract*—Auscultatory analysis using an electronic stethoscope has attracted increasing attention in the clinical diagnosis of respiratory diseases. Recently, neural networks have been applied to assist in respiratory sound classification with achievements. However, it remains challenging due to the scarcity of abnormal respiratory sound. In this paper, we propose a novel architecture, namely Waveform-Logmel audio neural networks (WLANN), which uses both waveform and log-mel spectrogram as the input features and uses Bidirectional Gated Recurrent Units (Bi-GRU) to context model the fused features. Experimental results of our WLANN applied to SPRSound respiratory dataset show that the proposed framework can effectively distinguish pathological respiratory sound classes, outperforming the previous studies, with 90.3% in sensitivity and 93.6% in total score. Our study demonstrates the high effectiveness of the WLANN in the diagnosis of respiratory diseases.

## I. Introduction

Respiratory diseases represent a leading cause of mortality worldwide, especially in children under five years old [1], imposing a substantial burden on a global scale. Research indicates that respiratory conditions originating in childhood often persist into adulthood, contributing to chronic health issues later in life [2]. Early identification, diagnosis, and management of pediatric respiratory diseases are therefore essential to prevent adverse outcomes such as respiratory failure and diminished lung function, ultimately improving long-term health and quality of life. Lung auscultation is a widely used diagnostic and treatment method for respiratory diseases, requiring only a stethoscope and no specialized equipment. This method involves analyzing respiratory sound characteristics, such as their presence and frequency, which are critical for the diagnosis, classification, and management of respiratory diseases and their exacerbations [3]. Despite its widespread use, traditional lung auscultation is constrained by notable limitations, particularly its reliance on the clinician's auditory skills and expertise. These limitations underscore the urgent need for the development and deployment of automated auscultation systems to enhance diagnostic accuracy and reliability.

Deep learning methods have proved to be effective in medical audio tasks, such as heart disease diagnosis, respiratory disease diagnosis, and speech disorders [4-6]. In recent years, pretrained models have made great contributions to audio classification task, especially Audio Spectrogram Transformer (AST), which splits a log-mel spectrogram into a sequence of image patches and extracts the attention information from patches using self-attention [7]. So far, fine-tuning AST has achieved state-of-the-art (SOTA) respiratory sound classification performances on open respiratory datasets [8,9]. Although these previous studies performed well on identifying normal respiration, they have only a moderate performance on detecting abnormal sounds [8,9].

Reliable identification of abnormal classes, especially crackles and wheezes, is pivotal for precise respiratory disease diagnosis. Crackles are discontinuous and low-pitched with non-musical auditory patterns linked to conditions like congestive heart failure, pneumonia, and lung fibrosis, while wheezes are continuous high-pitched tonal sequences observed in individuals with asthma. Since the log-mel spectrogram can only provide coarser resolution for higher frequencies, despite fine resolution for lower frequencies, and the AST output has low temporal resolution [10], the single spectrogram-AST architecture will result in a significant loss of frequency and temporal information of respiratory sound.

To tackle these issues, we here propose Waveform-Logmel audio neural networks (WLANN) with two major improvements as compared to the spectrogram-AST architecture. 1) A waveform-CNN system is designed to learn a time-frequency representation that possibly lacks in the spectrogram-AST system. 2) Bidirectional Gated Recurrent Units (Bi-GRU) are created to context model the fused features of waveform-CNN system and spectrogram-AST system, and learn the temporal context in the frame-level of respiratory sound. Specifically, the output of waveform-CNN system and the output of spectrogram-AST system are fused along the channel axis to obtain features from both the time-domain waveforms and log-mel spectrograms. The main contributions of this study are summarized as follows:

(1) We propose a model that autonomously extracts both the waveform and spectrogram features from the respiratory sound recording.

(2) The features fused from the waveform and spectrogram are fed to the Bi-GRU for context modeling in the frame-level.

This work was supported by the National Natural Science Foundation of China (62090030/62090031, 62274145), the National Key R&D Program of China (2021YFA1200502).

XXX-X-XXXX-XXXX-X/XX/$XX.00 ©20XX IEEE

(3) The proposed model shows superior performance on the classification of abnormal respiratory sounds, which is crucial for detection of respiratory diseases.

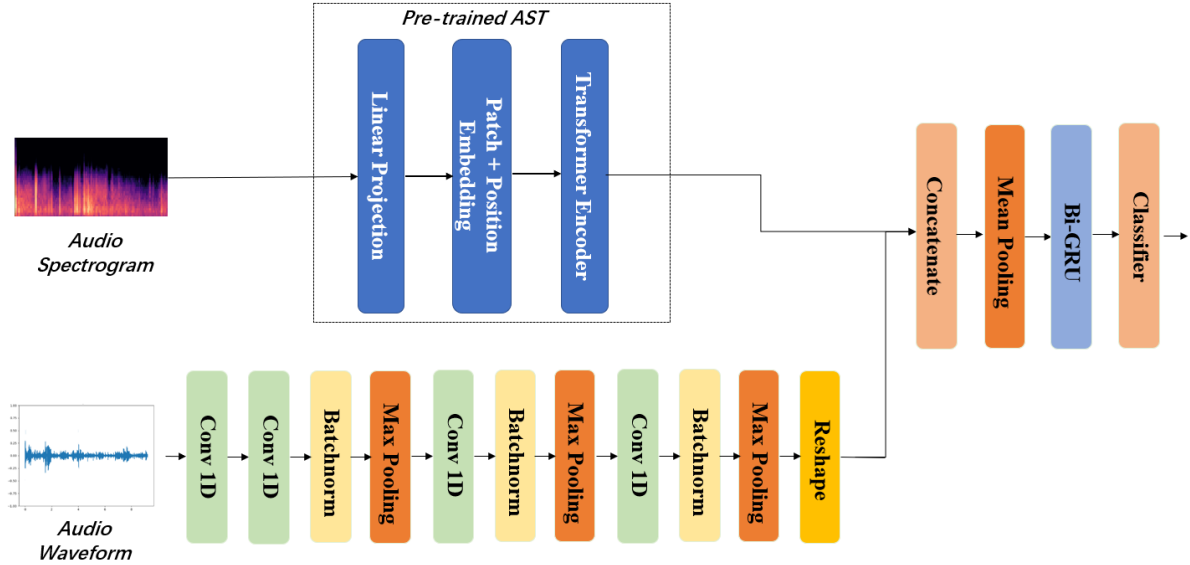

Fig. 1. The overall architecture of the Waveform-Logmel Audio Neural Networks. The proposed model uses both the waveform and log-mel spectrogram as the input feature and uses Bi-GRU to context model the fused features.

## II. RELATED WORK

Respiratory sound classification is an essential task of respiratory disease diagnosis, with the goal of identifying the type of respiratory sound in a lung auscultation clip. Recently, neural networks based methods have been successfully applied to predict the tags of respiratory, especially CNN-based methods [11,12]. Moreover, Convolutional Recurrent Neural Network (CRNN) architectures [13] can perform frame-level feature extraction and context modeling for respiratory sound classification. However, the lack of clinical abnormal breath sounds and unbalanced datasets limit the classification performance of CNN-based methods.

Another trend is to combine CNN with attention mechanism or purely based on attention mechanisms. Li et al [14] proposed a network that uses CNN to serialize the features of respiratory sounds, and the attention module to learn the serialized features, achieving promising results. Yuan et al [8] developed the purely attention-based AST, which are being widely used in the audio domain and achieved SOTA performance on respiratory sound classification [8,9].

## III. METHOD

The architecture of our proposed framework is illustrated in Figure 1. We integrate the outputs of the waveform-CNN system and the Spectrogram-AST system to create a novel representation. This approach allows us to leverage information from both time-domain waveforms and log-mel spectrograms effectively. The integration is performed along the channel dimension, where the waveform contributes additional information for audio classification, thereby complementing the log-mel spectrogram.

### A. Waveform-CNN System

The waveform-CNN is a time-domain respiratory sound classification system designed to process respiratory sound waveforms directly. The one-dimensional convolutional neural network (1D-CNN) serializes the waveform input and extracts time-domain features. Let the output of 1D-CNN be represented as $\mathbf{CO} \in \mathbb{R}^{T \times C}$, where $T$ denotes the number of frames and $C$ represents the number of channels. However, this output lacks critical frequency information, which plays a vital role in respiratory sound classification. To address this limitation, the output $\mathbf{CO}$ is reshaped to a tensor $\mathbf{WO} \in \mathbb{R}^{F \times T \times C/F}$, by splitting $C$ channels into $C/F$ groups, with each group containing $F$ frequency bins. This restructuring introduces frequency information into $\mathbf{WO}$, allowing it to represent features across both time and frequency domains. In this paper, the 1D-CNN architecture begins with an initial convolutional layer that employs a kernel size of 80 and a stride of 5, followed by three convolutional blocks. These blocks utilize the same kernel size but operate with a stride of 4. This configuration effectively corresponds to capturing 20 ms of respiratory audio per frame and producing 25 frames of feature representations per second, enabling fine-grained analysis of respiratory sounds.

### B. Spectrogram-AST System

The AST model has emerged as a pioneering architecture in the realm of spectrogram processing [15]. Structurally, the AST, similar to Vision Transformer [16], divides spectrogram into a sequence of overlapping or non-overlapping patches. Each patch is then linearly embedded into a fixed-size vector, allowing the transformer to process it as a sequence. Additionally, the AST model incorporates positional encodings, which are crucial in maintaining the order of the temporal sequence and a fundamental aspect of audio signals. Patches are fed to a transformer encoder and finally get the sound spectrogram representation $\mathbf{AO} \in \mathbb{R}^{F \times T \times C}$. Figure 2 illustrates the architecture of the employed AST. In this study, the respiratory recordings are resampled to 16 kHz and then converted into log-mel spectrograms. We split the spectrogram into a sequence of $16 \times 16$ patches with an overlap of 8 in both time and frequency dimension.

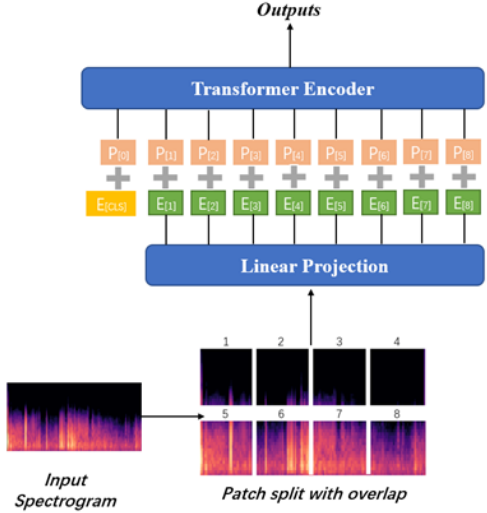

Fig. 2 The Audio Spectrogram Transformer (AST) architecture.

### C. Waveform-Logmel Audio Neural Networks

Considering the output of Waveform-CNN system $\mathbf{WO} \in \mathbb{R}^{F \times T \times C/F}$ and the output of Spectrogram-AST system $\mathbf{AO} \in \mathbb{R}^{F \times T \times C}$, we concatenate $\mathbf{WO}$ and $\mathbf{AO}$ along channel axis to form the final fused sequence $\mathbf{C} \in \mathbb{R}^{F \times T \times (C+C/F)}$, as illustrated in Figure 3. Similar to the Convolutional Recurrent Neural Network (CRNN) [17,18], $\mathbf{C}$ with temporal and frequency features is first aggregated (mean pooled) along the frequency axis to form a frame-level sequence, and then fed to the Bi-GRU for context modeling. A linear layer with sigmoid activation maps the output of the Bi-GRU to labels for classification.

### D. Waveform-Logmel Audio Neural Networks

Considering the imbalance of respiratory datasets, we employ the multi-class focal loss (MCFL) function [19] as expressed in the following equation:

$$\mathcal{L}_{MCFL} = -\sum_{i=1}^{C}(1-y_i)^\gamma p_i \log(y_i) \quad (1)$$

where $C$ denotes the number of categories, $p_i$ denotes a real probability distribution, and $y_i$ denotes a probability distribution of the prediction. A focusing parameter $\gamma$ is utilized to down-weighting normal respiratory contribution to the total loss so that the WLANN model focuses more on the features of abnormal respiratory sounds. We adopt the multi-class focal loss in our experiments with $\gamma$ =2.0.

## IV. EXPERIMENTS

### A. Dataset

We evaluate the performance of the proposed model on SPRSound respiratory dataset [20]. This dataset comprises 2,683 audio recordings and 9,089 annotated respiratory events collected from 292 patients aged between 1 month and 18 years. The recordings were obtained using a Yunting Model II Stethoscope with a sampling rate of 8 kHz and 16-bit precision. Expert clinicians carefully examined the recordings and annotated the onset (starting time) and offset (ending time) of each respiratory event. The dataset contains events categorized into seven distinct classes: Normal (N), Rhonchi (Rho), Wheeze (W), Stridor (Str), Wheeze and Crackle (Both), Coarse Crackle (CC), and Fine Crackle (FC). The duration of the audio recordings ranges from 0.3 to 15.4 seconds, while the individual respiratory events range from 0.1 to 7.2 seconds. Additionally, the dataset provides an officially defined split for training and testing, as shown in Table I. The first testing set (intra-patient) was derived from the same participants as the training set, while the second testing set (inter-patient) consists of data from different participants, allowing for a more robust evaluation of model generalizability.

TABLE I  THE DETAILED STATISTICS OF THE SPRSOUND DATASET.

| Type | Training | Testing-1 (intra-patient) | Testing-2 (inter-patient) | Total |
|---|---|---|---|---|
| **Normal** | 5159 | 688 | 1040 | 6887 |
| **Rhonchi** | 39 | 14 | 0 | 53 |
| **Wheeze** | 452 | 108 | 305 | 865 |
| **Stridor** | 15 | 2 | 0 | 17 |
| **Coarse Crackle** | 49 | 14 | 3 | 66 |
| **Fine Crackle** | 912 | 175 | 80 | 1167 |
| **Wheeze &Crackle** | 30 | 3 | 1 | 34 |
| **Total** | 6656 | 1004 | 1429 | 9089 |

### B. Data Pre-processing

After the original respiratory records are resampled to 16 kHz and filtered by a 4th-order Butterworth filter [21] with a passband of 40–850 Hz, we get the input waveforms for Waveform-CNN system. The passband range is chosen to maintain the fundamental frequencies of respiratory events, eliminate high-pitched artifacts, and reduce the effect of heart sounds [22,23]. The filtered waveforms are converted into a sequence of 128-dimensional log-Mel filterbank (fbank) features computed with a 25 ms Hamming window every 10 ms to generate log-mel spectrogram [7]. Moreover, data augmentation is utilized to improve the Spectrogram-AST system's ability to learn spectrogram representation while avoiding overfitting. We apply SpecAugment to augment data during training [24], which has been proposed for augmenting speech data for speech recognition. Time-warping and frequency masking are used for improving the robustness of Spectrogram-AST system to distortion of audio clips.

### C. Evaluation Metrics

The distribution of normal and abnormal samples in the SPRSound dataset [20] is highly imbalanced, which makes accuracy an unreliable metric for evaluating model performance. In contrast, sensitivity (SN) and specificity (SP) are more robust and reliable metrics in such scenarios [25]. Sensitivity measures the classifier's ability to correctly identify positive samples, while specificity quantifies the proportion of true negatives (correctly identified negative samples) out of all actual negatives. By considering both SN and SP, the evaluation achieves a more balanced assessment by focusing on the model's performance across individual classes. In addition to SN and SP, we also use average score (AS), harmonic score (HS), and total score (TS) to provide a comprehensive evaluation of our model's performance. The definitions and formulations of each metric are provided below:

$$\text{SN} = \frac{CAS}{TAS} \quad (2)$$

$$\text{SP} = \frac{CNS}{TNS} \quad (3)$$

$$\text{AS} = \frac{SN + SP}{2} \quad (4)$$

$$\text{HS} = \frac{2 \times SN \times SP}{SN + SP} \quad (5)$$

$$\text{TS} = \frac{AS + HS}{2} \quad (6)$$

where $CAS$ and $CNS$ refers to the number of correctly recognized abnormal sounds and normal sounds, respectively, while $TAS$ and $TNS$ refers to the number of total abnormal sounds and normal sounds.

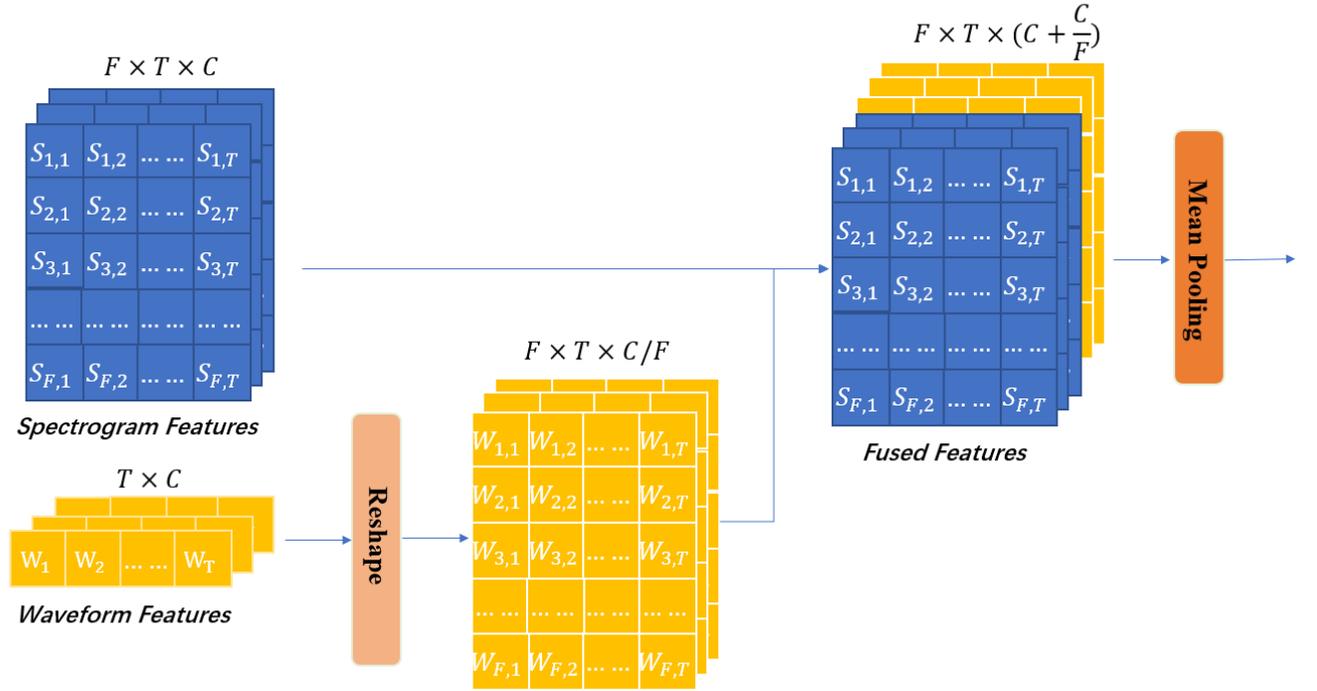

Fig. 3 Spectrogram features and waveform features are combined alone channel axis.

### D. Results

The results and comparison with these SOTA methods [26-30] are summarized in Table II. Compared with previous studies [30] that incorporates both log-mel spectrogram and raw time series, our model achieves state-of-the art results in SE and TS, with 90.3% in SE and 93.6% in total score, which demonstrates that our proposed architecture is more effective at learning input features and exhibits superior capability in distinguishing pathological respiratory sounds. Compared to the CNN or CNN+attention [26,30] architecture, our proposed model behaves a more balanced SE and SP, which is also significant for clinical breath sound diagnosis.

Furthermore, we have investigated different combinations of each neural networks on SPRSound respiratory dataset. As summarized in Table III, the combined input architecture outperforms the single input architecture, demonstrating that the incorporating of a rich set of features is crucial for enhancing the performance of the respiratory sound classification model. Additionally, the employment of a recurrent neural network (RNN) to process the respiratory sound features generated by the preceding network can obviously improve the model's sensitivity. This enhancement is attributed to the critical role of frame-level respiratory sound context features in distinguishing abnormal respiratory sounds.

TABLE II  OVERALL COMPARISON OF RESULTS ON SPRSOUND DATASET

| Method | Architecture | Sen (%) | Spe (%) | AS (%) | HS (%) | TS (%) |
|---|---|---|---|---|---|---|
| **Jun Li et al.[26]** | CNN | 79.4 | 84.7 | 82.1 | 82.0 | 82.0 |
| **Chen, Zizhao, et al. [27]** | ResNet18 | 67.7 | 93.9 | 80.8 | 78.7 | 79.7 |
| **Polymerized Feature [28]** | ML | 84.1 | 79.9 | 82.0 | 81.9 | 82.0 |
| **Ma, Wei-Bang, et al. [29]** | DenseNet169 | <u>88.7</u> | 93.1 | 90.9 | <u>90.9</u> | <u>90.9</u> |
| **TRespNET [30]** | CNN+attn | 84.0 | **98.0** | <u>91.0</u> | 90.0 | 90.0 |
| **WLANN (ours)** | AST+CNN+RNN | **90.3** | <u>96.9</u> | **93.6** | **93.5** | **93.6** |

TABLE III  RESULTS OF DIFFERENT COMBINATIONS OF EACH NEURAL NETWORK

| Method | Architecture | Sen (%) | Spe (%) | AS (%) | HS (%) | TS (%) |
|---|---|---|---|---|---|---|
| **Waveform-CNN** | CNN | 43.1 | 72.3 | 57.7 | 54.0 | 55.9 |
| **Waveform-CNN + Bi-GRU** | CNN+RNN | 44.2 | 76.5 | 60.4 | 56.0 | 58.2 |
| **Spectrogram-AST** | AST | 81.7 | 90.1 | 85.9 | 85.7 | 85.8 |
| **Spectrogram-AST+Bi-GRU** | AST+RNN | 85.3 | <u>96.6</u> | 91.0 | <u>91.0</u> | 91.0 |
| **Waveform-CNN+ Spectrogram-AST** | CNN+AST | <u>86.1</u> | 96.5 | <u>91.3</u> | 91.0 | <u>91.2</u> |
| **WLANN (ours)** | AST+CNN+RNN | **90.3** | **96.9** | **93.6** | **93.5** | **93.6** |

## V. CONCLUSION

In this study, we have demonstrated the effectiveness of our WLANN model, which combines Waveform-CNN system with Spectrogram-AST system, and achieves the SOTA results in sensitivity and total score. Using both waveform and log-mel spectrogram as the input feature can not only make up for the lack of frequency characteristics in the waveform, but also avoid the loss of features caused by information loss in the spectrogram. Furthermore, we have studied the effectiveness of recurrent neural networks (RNNs) in processing frame-level respiratory sound features, and found that it can strikingly enhance the model's sensitivity. Our proposed model exhibits outstanding performance in identifying abnormal sounds, advancing this research area closer to clinical applications.